\documentclass[preprint,onecolumn]{IEEEtran}

\usepackage{algorithm}
\usepackage{algorithmic}

\usepackage{graphics, graphicx}
\usepackage{subfig}
\usepackage{multirow}
\usepackage{caption}
\usepackage{physics}
\usepackage{xcolor}

\usepackage{bbm}
\usepackage{amssymb}
\title{Classical Simulation of Variational Quantum Classifiers using Tensor Rings}

\author{Dheeraj Peddireddy, Vipul Bansal, and Vaneet Aggarwal\thanks{D. Peddireddy and V. Aggarwal are with Purdue University, West Lafayette IN 47907, email: \{dpeddire, vaneet\}@purdue.edu. V. Bansal is with Indian Institute of Technology Roorkee, India. }}

\if 0
\author[inst1]{Dheeraj Peddireddy}
\author[inst2]{Vipul Bansal}
\author[inst1]{Vaneet Aggarwal}

\affiliation[inst1]{organization={School of Industrial Engineering, Purdue University},%Department and Organization
            city={West Lafayette},
            postcode={47907}, 
            state={IN},
            country={USA}}
\affiliation[inst2]{organization={Indian Institute of Technology Roorkee},%Department and Organization
            city={Roorkee},
            country={India}}
\fi 

\usepackage{setspace}
\begin{document}
\maketitle

\begin{abstract}
In recent times, Variational Quantum Circuits (VQC) have been widely adopted to different tasks in machine learning such as Combinatorial Optimization and Supervised Learning. With the growing interest, it is pertinent to study the boundaries of the classical simulation of VQCs to effectively benchmark the algorithms. Classically simulating VQCs can also provide the quantum algorithms with a better initialization reducing the amount of quantum resources needed to train the algorithm. This manuscript proposes an algorithm that compresses the quantum state within a circuit using a tensor ring representation which allows  for the implementation of VQC based algorithms on a classical simulator at a fraction of the usual storage and computational complexity. Using the tensor ring approximation of the input quantum state, we propose a method that applies the parametrized unitary operations while retaining the low-rank structure of the tensor ring corresponding to the transformed quantum state, providing an exponential improvement of storage and computational time in the number of qubits and layers. This approximation is used to implement the tensor ring VQC for the task of supervised learning on Iris and MNIST datasets to demonstrate the comparable performance as that of the implementations from classical simulator using Matrix Product States.
\end{abstract}
%\begin{frontmatter}



\if 0
\begin{highlights}
\item 
\item Research highlight 2
\end{highlights}
\fi


%\end{frontmatter}


\section{Introduction}
Quantum computing has been demonstrably proven to be superior than classical computing when solving problems such as searching unstructured databases \cite{10.1145/237814.237866}  or factorization of large numbers \cite{shor1999polynomial}. Due to the rising interest in the computational capabilities of noisy intermediate scale quantum (NISQ) computers and the availability of experimentation platforms, there has been a rapid growth in the development of circuit based algorithms. VQCs or Parametrized Quantum circuits are one such class of the circuit based algorithms that has been studied extensively \cite{cerezo2021variational}, specifically in context of their capabilities in solving various combinatorial optimization problems and intrinsic energy problems of molecules which were either intractable or computationally expensive using classical devices. These studies have also been extended to use VQCs as a replacement for artificial neural networks (ANN) in discriminative and generative tasks. Despite its infancy, there are numerous works exploring the algorithmic aspects and applications of VQCs. Several works have developed Quantum Approximate Optimization Algorithm and its generalizations to train parametrized quantum circuits using classical optimizers \cite{zhou2020quantum, farhi2014quantum,mcclean2016theory}. VQCs in the context of supervised learning, have first introduced in Mitarai et al. \cite{mitarai2018quantum} which theoretically and numerically proved that quantum circuits can approximate non-linear functions similar to neural networks. Circuit based algorithms have been developed to emulate several methods from the classical machine learning literature such as Long Short-Term Memory \cite{chen2020quantum} and Convolutional Networks \cite{cong2019quantum,liu2021hybrid}. It has also been shown that that the VQCs have a better expressive power compared to neural networks \cite{Du_2020}. 

\begin{figure*}[!t]
 \centering
 \includegraphics[height=7cm, width=14cm]{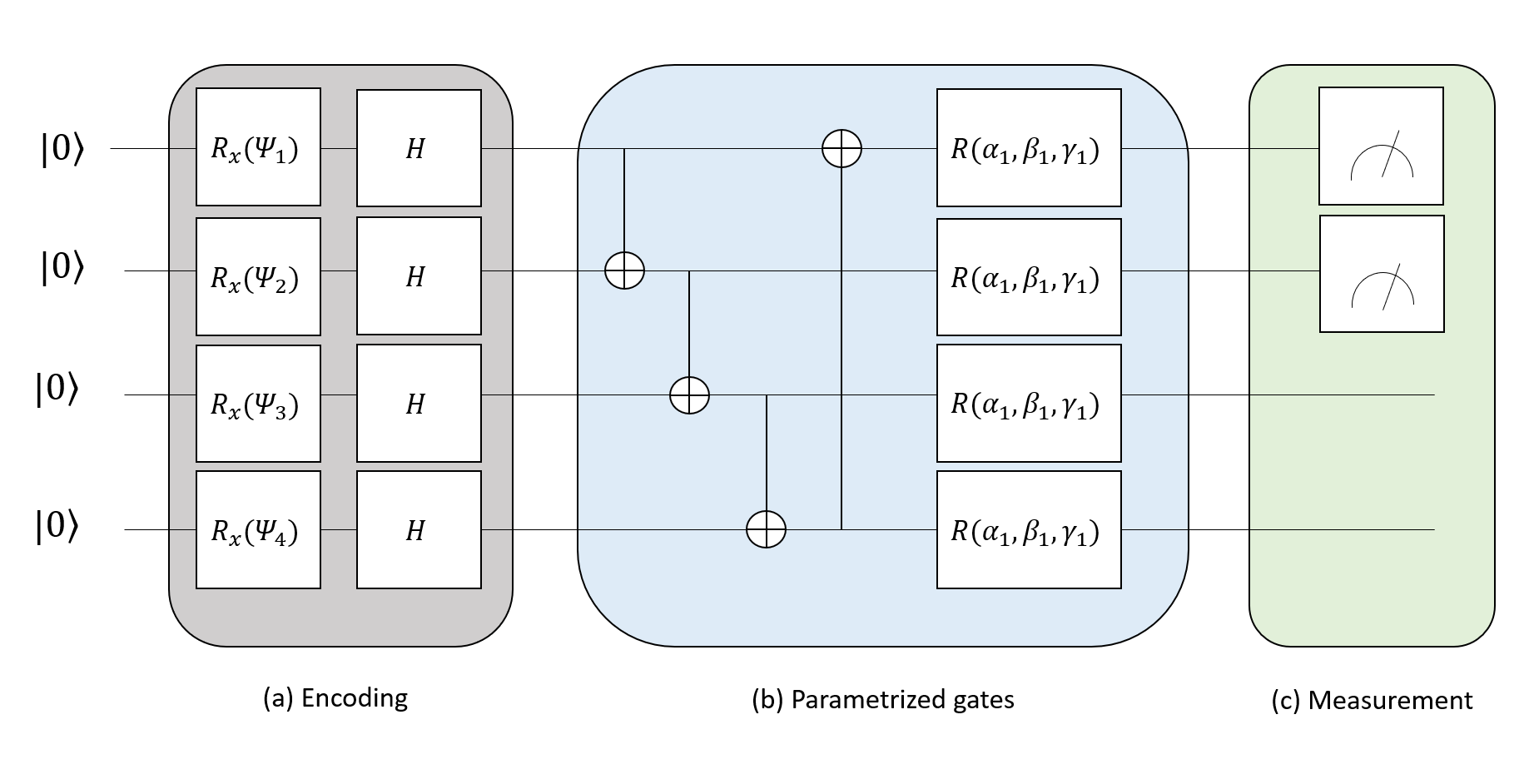}
 \caption{\textbf{Illustration of a generic 4-qubit variational quantum circuit} $R_x(\cdot)$ gates denote a rotation about the x-axis on Bloch sphere by a certain angle. H represents the Hadamard gate. (a) The first part of the circuit encodes the classical features $\psi$s onto the qubits, (b) the rotation gates with free parameters that transform the qubits are denoted by $R(\alpha, \beta, \gamma)$ and (c) a subset of the qubits are measured to obtain the output}
 \label{fig:genvqc}
\end{figure*}

Despite several studies indicating the promise of quantum superiority in machine learning, a key challenge in the current era of quantum computing is the limited quantum resources which in turn limits the number of qubits and the circuit depth. The other challenge is the inherent tendency of quantum devices towards decoherence, random gate errors and measurement errors which might severely limit the training of the variational circuits on quantum simulators. A possible solution to address these issues is to implement the training classically before deploying them on quantum simulators in order to avoid the compounding effects of decoherence throughout the numerous iterations usually required for learning. However, the simulation of quantum computations on a classical device grows exponentially harder in both the number of qubits and the circuit depth or the number of layers. Therefore it is pertinent to study the scope of classical simulation of quantum systems in the realm of learning not only to assist in the training, but also to validate the results from the quantum devices. Moreover, the much sought after quantum advantage is based on the assumption that there exists some tasks which can be solved much easily using the quantum devices over classical devices, so it is imperative to study the boundaries of classical simulation. One of the well studied approaches to compress the information in the quantum state is using the matrix product states (MPS)  \cite{schollwock2011density}. A recent work from Zhou et al.\cite{PhysRevX.10.041038} has constructed a truncation technique that simulates a real quantum device by inducing a noise into the MPS approximation of a quantum state. This work extends upon the previous literature, by considering a noisy tensor ring approximation to train VQCs for supervised learning over popular datasets.

We note that the MPS (tensor train or TT) has been used extensively for scalable simulation of quantum circuits. However, such a decomposition suffers from the following limitations: (i) TT model requires rank-1 constraints to
the border factors i.e., TT ranks are typically small for near-border factors and large for the middle factors, and (ii) the multiplications of the TT factors are not permutation invariant. In order to alleviate these issues, researchers have started to use cyclic MPS or  tensor ring (TR) representation in the classical machine learning \cite{wang2017efficient,Wang_2018_CVPR,malik2021sampling}. TR decomposition removes the unit rank constraints for the boundary tensor factors and utilizes a trace operation in the decomposition. The multilinear products between factors also have no strict ordering and the factors can be circularly shifted \cite{wang2017efficient} allowing for all ranks to be the same. TR representation has been shown to significantly outperform TT representation for data completion \cite{wang2017efficient}, compression of classical neural networks \cite{Wang_2018_CVPR}, etc. These properties allow for a better approximation of circular entanglement in quantum circuits motivating the use of TR representation for compressing the quantum state in the VQC in this paper. This manuscript proposes a classical algorithm, Tensor Ring Variational Quantum Circuit (TRVQC) which represents the quantum state using a low-rank tensor ring and the transformations corresponding to the single and two-qubit gates in the VQC are applied to the TR while retaining its structure. In a naive Tensor Network approximation, two-qubit tranformations do not preserve the low rank of a tensor ring which TRVQCs address by approximating the resultant high rank tensor using truncated singular value decomposition.

\begin{figure*}[!t]
 \centering
 \includegraphics[keepaspectratio ,width=0.8\textwidth]{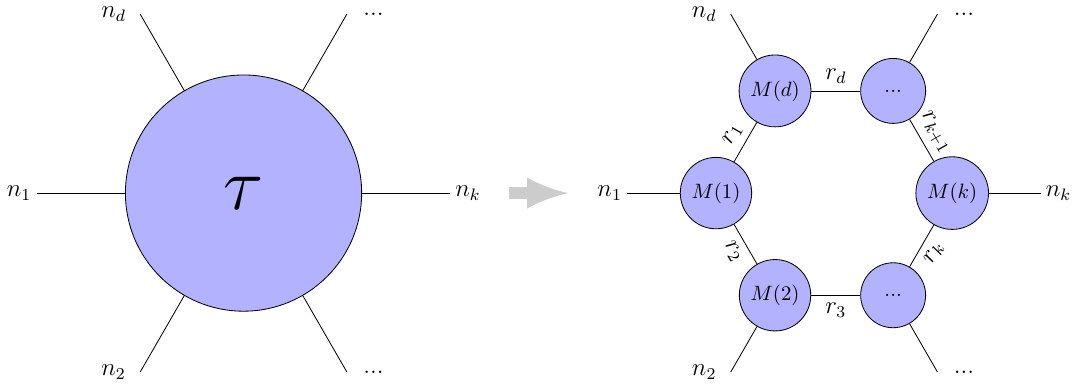}
 \caption{Illustration of tensor ring decomposition of a rank-d tensor}
 \label{fig:trd}
\end{figure*}

%This is done by introducing a finite error rate $\epsilon$ when computing a two qubit gate. It is shown that the computation time grows linearly in the number of qubits and the circuit depth which is represented by the number of two qubit gate layers in the circuit. The technique was implemented with 60 qubits and a circuit depth of 200 which only takes a small fraction of time to simulate compared to the exact simulation. 

%Building up on the work in \cite{PhysRevX.10.041038}, we propose a noisy tensor ring parametrization to train the variational quantum circuit. This technqiue addresses a major shortcoming i.e. scalability of the variational quantum circuits, briding the gap between large scale machine learning and quantum computing. 

To the best of authors' knowledge, this is the first work on using scalable TR based VQCs for classification. We note that the approximation of two-qubit gates using singular value decomposition brings non-linearity in the operations, and is inherently similar to  adding the ReLU or soft-max operations in classical neural networks. We believe that this non-linearity further helps in the selection of the learnable parameters thus helping achieve improved performance. Furthermore, VQCs have been popularized for their robustness to noise making them well-suited for NISQ era computers indicating that TRVQCs can share a similar robustness to the noise added through truncation.

The rest of the manuscript is organized as follows: Section \ref{sec:2} discusses the related literature corresponding to the approximation methods and their utility in QML. Section \ref{sec:3.1} discusses the workings and general architecture of VQCs. Section \ref{sec:3.2} introduces the mathematical notion of the TR approximation of a quantum state and the transformation of the TR with respect to the single and two qubit rotations. Section \ref{sec:3.3} describes the supervised learning algorithm that utilizes the TR approximation and transformation techniques from Section \ref{sec:3.2}. Section \ref{sec:4} demonstrates the results from the experiments comparing the proposed architecture with a Matrix Product State simulator accompanied by a discussion on the storage and computational complexity. Finally, Section \ref{sec:5} concludes the manuscript by briefly summarizing the results from the work and discussing limitations and future directions of research.

%We evaluate the performance of the proposed approach on Iris and MNIST datasets. In case of the MNIST dataset, the performance of the proposed method is better than the implementation of VQCs in Qiskit's classical simulator \cite{cross2018ibm} which uses the full MPS state. This demonstrates that the proposed architecture using TR representation outperforms not using approximation. This is likely because of introduction of non-linearity and efficient computation of gradients in the compressed TR representation using established frameworks like PyTorch \cite{torch}. Further, due to linear scaling of computation and storage, this structure allows for scaling in the number of qubits and circuit depth, which has the potential to demonstrate quantum supremacy or lack thereof in machine learning. 

\section{Related work}\label{sec:2}
Over the years, MPS and its generalizations have been used extensively in the classical simulation of quantum circuits \cite{vidal2003efficient, markov2008simulating} owing to their capability of accurately capturing low to moderate entanglements in many-qubit quantum states. Several approximate techniques have been developed using the MPS parametrization such as Density Matrix Renormalization Group (DMRG) \cite{white1992density}, Projected entangled pair states (PEPS) \cite{verstraete2008matrix} and multi-scale entanglement renormalization ansatz (MERA) \cite{vidal2007entanglement}. Qiskit, a popular quantum programming platform builds upon the work proposed in Vidal et al. \cite{vidal2003efficient} to implement quantum circuits by simulating the quantum state using Matrix Product States. However using this method, tensor size can grow exponentially when there are a large number of two qubit gates in the circuit. To address this issue,  Zhou et al. \cite{PhysRevX.10.041038} proposed a truncation based on the singular values of the entangled tensors, which effectively represent the decoherence which limits the amount of entanglement that can be built into a quantum state in real quantum devices. This work was demonstrated to have a linear complexity in worst case when evaluating expected values from a quantum circuit at a small cost to accuracy. However, this technique was not evaluated on VQCs which are robust to the noise.

 \begin{figure*}[!t]
 \centering
 \includegraphics[height=7cm, width=13cm]{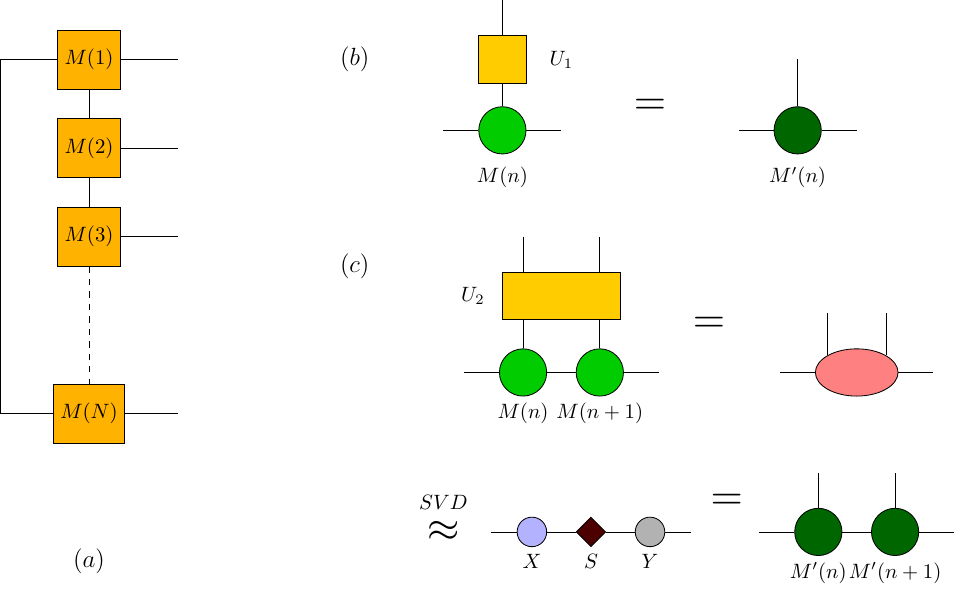}
 \caption{(a) A tensor ring structure representing the N-qubit state with bond dimension $\chi$ (b) Applying a single qubit gate to the n-th qubit of a tensor ring (c) Applying a two qubit gate to n-th and n+1-th qubits of a tensor ring and using the truncated SVD to decompose the resultant matrix into new tensors}
 \label{fig:tn}
\end{figure*}

In the context of machine learning, MPS along with Tree Tensor Network approximation of quantum circuits have been successfully demonstrated in performing binary classification \cite{huggins2019towards}. PEPS-based \cite{cheng2021supervised} and MPS-based \cite{efthymiou2019tensornetwork} frameworks have been used for the task of image classification which utilizes the automatic gradients from the open source libraries. Although promising in cases of low entanglement, these works still suffer from the same problem as in Vidal et al. \cite{vidal2003efficient}. A quantum-inspired framework is proposed in Stoudenmire et al. \cite{stoudenmire2016supervised} which uses tensor networks to represent the weight matrix in classical supervised learning models and train the tensor network directly. This work uses a singular value truncation method similar to the one in Zhou et al. \cite{PhysRevX.10.041038} but the network architectures were developed independent of VQCs and do not involve quantum gates which are a crucial aspect of VQC-based learning algorithms. To address this shortcoming, we propose a method that utilizes a similar truncation method to retain the low-rank tensor ring as the unitary transformations from the VQC are applied to the TR representation of the quantum state.

\section{Methodology}\label{sec:3}

\subsection{Variational Quantum Circuits}\label{sec:3.1}

VQCs are quantum circuits with unitary rotation gates parametrized by trainable parameters. Several quantum algorithms have been developed to train the aforementioned VQCs for tasks like binary optimization, approximation, and classification. These variational quantum algorithms are usually quantum-classical hybrid algorithms where the loss function is evaluated using unitary transformations on quantum simulators and the parameters of the circuit are trained using classical optimization algorithms. VQCs can be used in similar applications as that of neural networks which primarily include function approximation. Quantum circuits use two qubit gates (usually controlled rotations) to encode entanglement into the quantum state which acts analogous to the activation functions in neural networks. Note that the  quantum circuits (excluding measurement) are linear reversible transformations and multitude of classical machine learning literature has established the significance of non-linearity in function approximation. This will be one of the key changes in our approximation compared to the previous studies, which will bring in non-linearity into the approximation.\newline 
We  describe the workings of a general VQC along with a pictorial representation of an example in Figure \ref{fig:genvqc}. Initially, the classical data is encoded into the qubits using rotations parametrized by the input data. The prepared qubit state is then transformed through a series of unitary rotation parametrized by trainable weights and entanglements. A subset of the transformed qubits are then measured to obtain the output in the form of expected value. The expected values are decoded into the appropriate class labels. A loss is computed akin to the neural networks using the decoded output and the true labels from the dataset. The free parameters are then updated using classical optimization methods like gradient descent, Adam, BFGS, etc. The trained circuit can then be used to predict the labels of the test data.
 \begin{figure*}[!t]
 \centering
 \includegraphics[height=6cm, width=16cm]{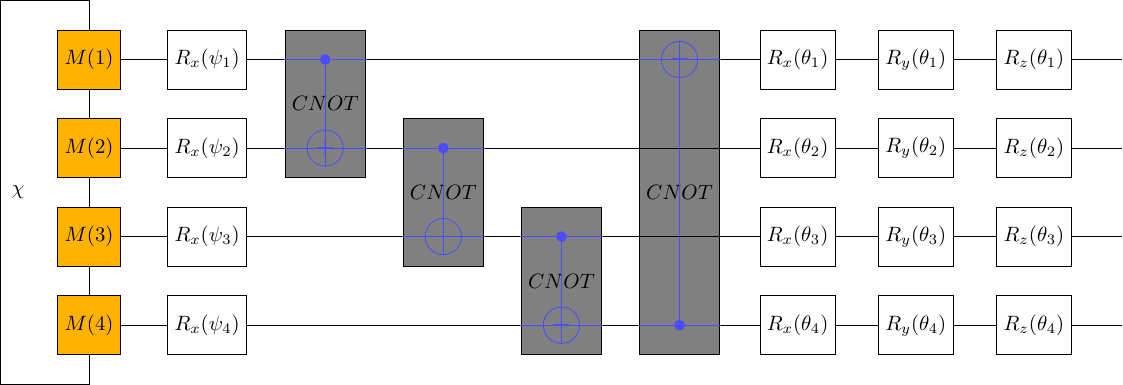}
 \caption{Circuit architecture for the Iris Classification problem}
 \label{fig:expc}
\end{figure*}

\subsection{ Tensor Ring Parametrization of VQC}\label{sec:3.2}
 Quantum circuits  become too difficult to simulate on classical computers for larger number of qubits since the storage and computation is exponential in the number of qubits.  Tensor networks \cite{orus2014practical} can be thought of as a graphical representation of tensors, where each node in the graph is a tensor with a finite rank (a rank-0 tensor is a scalar, rank-1 tensor is a vector, rank-2 tensor is a matrix etc.). A tensor ring is a tensor network which aims to represent a higher order tensor by a sequence of 3rd order tensors that are multiplied circularly as shown in Figure \ref{fig:trd}. A higher order tensor $\tau$ of rank $d$ is decomposed into $d$ tensors denoted by $M(k)$ each of rank 3 spanned by the indices $\{ n_k, r_k, r_{k+1} \}$ where $r_k$ and $r_{k+1}$ represent the bond dimension (or elements of the rank vector).

 It has been shown that the tensor rings are a better and more effective compression algorithm than tensor trains (or MPS format) \cite{wang2017efficient,Wang_2018_CVPR}, which motivates our choice. TT representation has lower border ranks and larger interior ranks which limit an efficient representation, while all ranks could be made similar in tensor ring format and thus decreased ranks helps achieve better representation at the same number of parameters. We also note that the many ansatz or architectures of VQCs have a cyclic entanglement, where the CNOT from the last qubit to the first qubit will not be directly possible using the MPS representation, which further motivates the use of tensor ring format in this paper. %We consider an $N$ qubit circuit where two qubit gates can only be applied to the adjacent qubits or the first and the last qubit. 
We represent the quantum state $\ket{\phi}$ using $N$ tensors corresponding to $N$ qubits with the $n$-th tensor denoted by $M(n)$, as below:

% It is also to be noted that in \cite{PhysRevX.10.041038}, the ansatz are restricted to two qubit gates applied only to consecutive qubits which is the same as in the proposed method except the tensor ring structure allows one additional connection with the first and last qubits .

\begin{equation}\label{eq:one}
\centering
\begin{aligned}
\ket{\phi} =  & \sum_{i_1 \ldots i_N}\sum_{\mu_1 \ldots \mu_N} M(1)_{\mu_N \mu_1}^{i_1} M(2)_{\mu_1 \mu_2}^{i_2} \ldots \\
& \space \ldots M(N)_{\mu_N \mu_1}^{i_N} \ket{i_1 i_2 \ldots i_N} 
\end{aligned}
\end{equation}
where $i_n \in \{ 0, 1 \}$ are the physical indices spanning the $2^N$ dimensional Hilbert space while $\mu_n \in \{ 1, \ldots, \chi_n \}$ are the bond indices of the tensors which control the maximum amount of entanglement captured by the tensor ring. If $\chi_n$ is allowed to grow large, the tensor ring would be able to capture the exact quantum state information at a computational cost. When designing the training algorithm, $\chi_n$ is considered to be one of the hyperparameter choices along with batch size, learning rate, etc. We assume $\chi_n=\chi$ for all $n$, reducing the number of hyperparameters following Wang et al. \cite{wang2017efficient}. This bond-dimension $\chi$ is called the tensor ring rank.  A tensor ring parametrization of an N-qubit state is illustrated in Figure \ref{fig:tn}(a). Each square in the ring is a tensor of dimension $\chi \times \chi \times 2$. The open bonds of the tensor ring index the 2-dimensional Hilbert space of the qubit corresponding to the site. We explain the steps to apply different kinds of unitary operations on the tensor ring approximation of a quantum state in the following paragraphs.

We first consider a one qubit rotation which is  a tensor contraction of a $2 \times 2$ matrix with the tensor corresponding to the qubit. A one qubit rotation (represented by a unitary matrix $U$) on  $n^{th}$ qubit is given by the following equation:

\begin{equation}\label{eq:two}
M'(n)_{\mu_{n-1} \mu_n}^{i'_n} = \sum_{i_n}U_{i'_n i_n} M(n)_{\mu_{n-1} \mu_n}^{i_n}
\end{equation}

In order to perform a two qubit gate transformation on qubits $n$ and $n+1$, we first transform the tensor ring into an orthogonal form centered around the qubits of interest. A series of operations are performed followed by a    singular value decomposition as shown below. The two tensors at $n$ and $n+1$ are first contracted along the shared bond index which is given by:

\begin{equation}\label{eq:three}
T_{\mu_{n-1} \mu_{n+1}}^{i_n i_{n+1}} = \sum_{\mu_n} M(n)_{\mu_n-1 \mu_n}^{i_n} M(n+1)_{\mu_n \mu_{n+1}}^{i_{n+1}}
\end{equation}

The two qubit gate ($U$, reshaped into $U_{i'_n i'_{n+1} i_n i_{n+1}}$) is then applied on the two qubit tensor computed in the previous equation:

\begin{equation}\label{eq:four}
(T')_{\mu_{n-1} \mu_{n+1}}^{i_n i_{n+1}} = \sum_{i_n i_{n+1}} U_{i'_n i'_{n+1} i_n i_{n+1}}  T_{\mu_{n-1} \mu_{n+1}}^{i_n i_{n+1}}
\end{equation}

Finally, we reshape the tensor $T'$ into a matrix of shape $(i'_n \times \mu_{n-1}) \times (i'_{n+1} \times \mu_{n+1})$, and perform singular value decomposition of the matrix:

\begin{equation}\label{eq:five}
(T')_{\mu_{n-1} \mu_{n+1}}^{i_n i_{n+1}} = \sum_{\mu_n} X_{\mu_{n-1} \mu_n}^{i'_n} S_{\mu_n} Y_{\mu_{n} \mu_{n+1}}^{i'_{n+1}}
\end{equation}

where $X$ and $Y$ matrices are composed of the orthogonal vectors and $S_\mu$ contains the singular values of the matrix $T'$. The matrix has $2 \chi$ singular values irrespective of the two qubit gate structure where $\chi$ denotes the bond dimension of the tensor ring. We then truncate the $S_{\mu}$ matrix to keep only the $\chi$ largest singular values and the resulting matrix is denoted by $S'_{\mu}$ . $X$ and $Y$ are truncated to only keep the orthogonal vectors corresponding to the $\chi$ largest singular values. The new tensors corresponding the two qubits are given by:

\begin{equation}\label{eq:six}
M'(n)_{\mu_{n-1} \mu_n}^{i_n} = X_{\mu_{n-1} \mu_n}^{i_n} S'_{\mu_n}
\end{equation}

\begin{equation}\label{eq:seven}
M'(n+1)_{\mu_{n} \mu_{n+1}}^{i_{n+1}} =  Y_{\mu_n \mu_{n+1}}^{i_{n+1}}
\end{equation}

Without the approximation, the operation of 2-qubit gate will be exponential in the number of qubits, the above operations are $O(1)$ in the number of qubits, and thus helps scalability of the approach. The single qubit and two qubit transformations are demonstrated in Figure \ref{fig:tn}.

\begin{algorithm}[!h]
\caption{Training TRVQC on a single batch}
\textbf{Input}: Input data , ground truth, VQC ansatz ($\mathbbm{U}$)\\
\textbf{Parameters}: Weight vector ($\theta$) corresponding to the variational layers, Optimization hyperparameters \\
\textbf{Output}: Updated weight vector
\begin{algorithmic}[1]
    \FOR{each sample in batch}
    \STATE Compute the tensor ring decomposition corresponding to the initial state $\ket{0}^{\otimes N}$ which is a series of $\chi \times \chi \times 2$ tensors with only the element at index (1,1,1) as 1 and rest of the elements as zeros
    \FOR{each gate $U$ in $\mathbbm{U}$}
    \IF{$U \in \mathbbm{C}^{2\times2}$}
    \STATE Transform the TR following the procedure from the equation \ref{eq:two}
    \ELSIF{$U \in \mathbbm{C}^{4\times4}$}
    \STATE Transform the TR following the procedure from equations \ref{eq:three}-\ref{eq:seven}
    \ENDIF
    \ENDFOR
    \STATE Contract the transformed tensor ring to obtain the full transformed quantum state
    \STATE Choose bitstrings corresponding to the measurements to assign to the classification labels
    \STATE Compute sigmoid probabilites of each label from the amplitudes of the corresponding bitstrings
    \STATE Compute cross-entropy loss using ground truth and the probabilites
    \ENDFOR
    \STATE Aggregate loss over batch size
    \STATE Compute the gradients of the cross-entropy loss with respect to weights of the variational layers
    \STATE Update the weights using classical gradient-based optimization methods
\end{algorithmic}
 \label{algo:one}
\end{algorithm}

\subsection{Training}\label{sec:3.3}

\begin{table*}[!h]
\centering
\resizebox{0.7\textwidth}{!}{
\begin{tabular}{|c|c|ccc|ccc|}
\hline
Dataset                                                                  & \#qubits & \multicolumn{3}{c|}{TR-VQC} & \multicolumn{3}{c|}{MPS-VQC} \\ \hline
                                                                         &          & 1 layer & 2 layer & 3 layer & 1 layer  & 2 layer & 3 layer \\ \hline
Iris                                                                     & 4        & 67.37   & 80.53   & 82.63   & 73.16    & 75.79   & 83.68   \\ \hline
\multirow{2}{*}{\begin{tabular}[c]{@{}c@{}}reduced\\ MNIST\end{tabular}} & 4        & 81.67   & 83.73   & 77.97   & 79.66    & 81.02   & 77.63   \\ \cline{3-8} 
                                                                         & 8        & 75.93   & 81.69   & 80.02   & 57.96    & 65.42   & 67.11   \\ \hline
\end{tabular}}
\caption{Test accuracy from training VQCs using the proposed tensor ring method (denoted by TR-VQC) benchmarked against naive Matrix Product State approximation (denoted by MPS-VQC) with varying number of qubits and circuit ansatz on Iris and reduced MNIST datasets}\label{tab:results}
\end{table*}
 
We first explain the overall VQC architecture followed by the training procedure. We prepare the initial $N$ qubit state as $\ket{0}^{\otimes N}$, which is then converted to a TR representation as $M(i)$ which is a $\chi\times\chi\times2$ tensor with only the element indexed $(1,1,1)$ as $1$ and rest of the elements as zeros. Then, the data features are used to perform single qubit rotations as illustrated in the {\bf encoding} step of Fig. \ref{fig:genvqc} (as one possible approach, while any other encoding schema can be used).   The {\bf parametrized gate circuit} is chosen as a modification of that illustrated in Fig. \ref{fig:genvqc}. This gate circuit is repeated $d$ times to illustrate the depth of the circuit. Finally, some or all of the qubits are {\bf measured}, based on which the loss function is determined for training. We evaluate the ``forward pass'' of the circuit by computing the single qubit and two qubit transformations with respect to the input tensor ring as discussed in the previous sub-sections to obtain the final output obtained by measuring the final state. The loss is then computed using both the final output and the ground truth, and the gradients are calculated by backpropagating through the tensor ring structure similar to neural networks. For high dimensional data like images, it is not uncommon to use Principal Component Analysis, Autoencoders, etc. to reduce the dimensionality of the data. In such cases, given that the tensor network algebra can be efficiently implemented using the popular machine learning libraries, it is possible to train the algorithm end-to-end, i.e., both the the parameters of the VQC and dimensionality reduction layer. This is not the case in most QML architectures where the preprocessing layer has to be pretrained. The details of backpropagation through quantum gates and singular value decomposition can be found in Watabe et al. \cite{watabe2019quantum} and Li et al. \cite{li2009efficient}. Given the compression obtained by the tensor ring representation and the the SVD used at the two qubit gates, the training of the proposed method would be much faster than training the full VQC simulated classically. Once trained, the parameters can be deployed on a real quantum device to evaluate the performance on unseen data. The steps involved in training the TRVQC on a single batch are listed in Algorithm \ref{algo:one}. For the simplicity of notation we represent the VQC ansatz as an ordered set of one qubit and two qubit gates that are to be applied to the input state, represented by $\mathbbm{U}$. The algorithm is repeated over multiple randomized batches for improved results.

\section{Evaluations}\label{sec:4}
To illustrate the capabilities of the proposed method for simulating VQCs, we perform experiments using various circuits with different number of qubits and layers on two datasets; Iris and a subset of MNIST containing classes 3 and 7. We also compare these results against the same circuits implemented on Qiskit's classical simulator using full Matrix Product States. As discussed before, Qiskit's simulator uses the work from Vidal et al. \cite{vidal2003efficient} which can grow exponentially in cases of high entanglement i.e. large number of two-qubit gates which is increasingly possible with more number of qubits and layers. Tensor ring rank $\chi$ is set to $8$ for all experiments unless otherwise specified. It is to be noted that optimizing the circuit ansatze for performance as discussed in Du et al. \cite{du2020quantum} is left out of the scope of the simulations. All the results are averaged over 5 runs of training with different initialization.

\subsection{Iris Classification}\label{sec:4.1}

 The Iris data set includes four features and three labels. The dataset is split in a ratio of $3:1$ for training/testing data. The circuit ansatz chosen for the problem is illustrated in Figure \ref{fig:expc}. The features are normalized before being encoded into the qubits. The initially prepared 4-qubit state of $\ket{0000}$ is decomposed into a tensor ring as described before. The features $\psi$ extracted from the data are encoded onto the qubits using a layer of $R_x$ gates. This is followed by a sequence of CNOT gates entangling all the consecutive qubits. After this, a sequence of single-qubit gates, i.e., $R_x$, $R_y$, and $R_z$ , are applied to each qubit with trainable parameters. Here, $R_x(\cdot), R_y(\cdot), R_z(\cdot)$ are rotation gates about the x, y and z axes respectively on the Bloch sphere. The CNOT gate is represented by a connection between two qubit ``wires" as shown in Figure \ref{fig:expc}. The circular end of the connection represents the target qubit and the dot end represent the control qubit. Out of the $2^4$ possible measurements obtained from the VQC, we select amplitudes of 3 measurements (bitstrings) for representing the 3 classes of the dataset. We further apply softmax function to selected measures to convert the expected values to class probabilities. Then, we utilize a cross entropy loss and back-propagate to obtain gradients for each trainable parameter. In this experiment, we use an Adam optimizer with a learning rate of $0.01$ for $50$ epochs for a batch size of $4$ . The simulations are repeated with three circuits containing 1, 2 and 3 parametrized layers. The results of the experiments along with the results from the Qiskit simulations are listed in Table \ref{tab:results}. 

% It has been shown that the cost of applying a two qubit gate dominates the SVD decomposition \cite{PhysRevX.10.041038} and hence the overall evaluation of the circuit is much more efficient than the exact simulation. 

\begin{figure*}[!t]
\centering
	\subfloat{\includegraphics[keepaspectratio,width=0.5\textwidth]{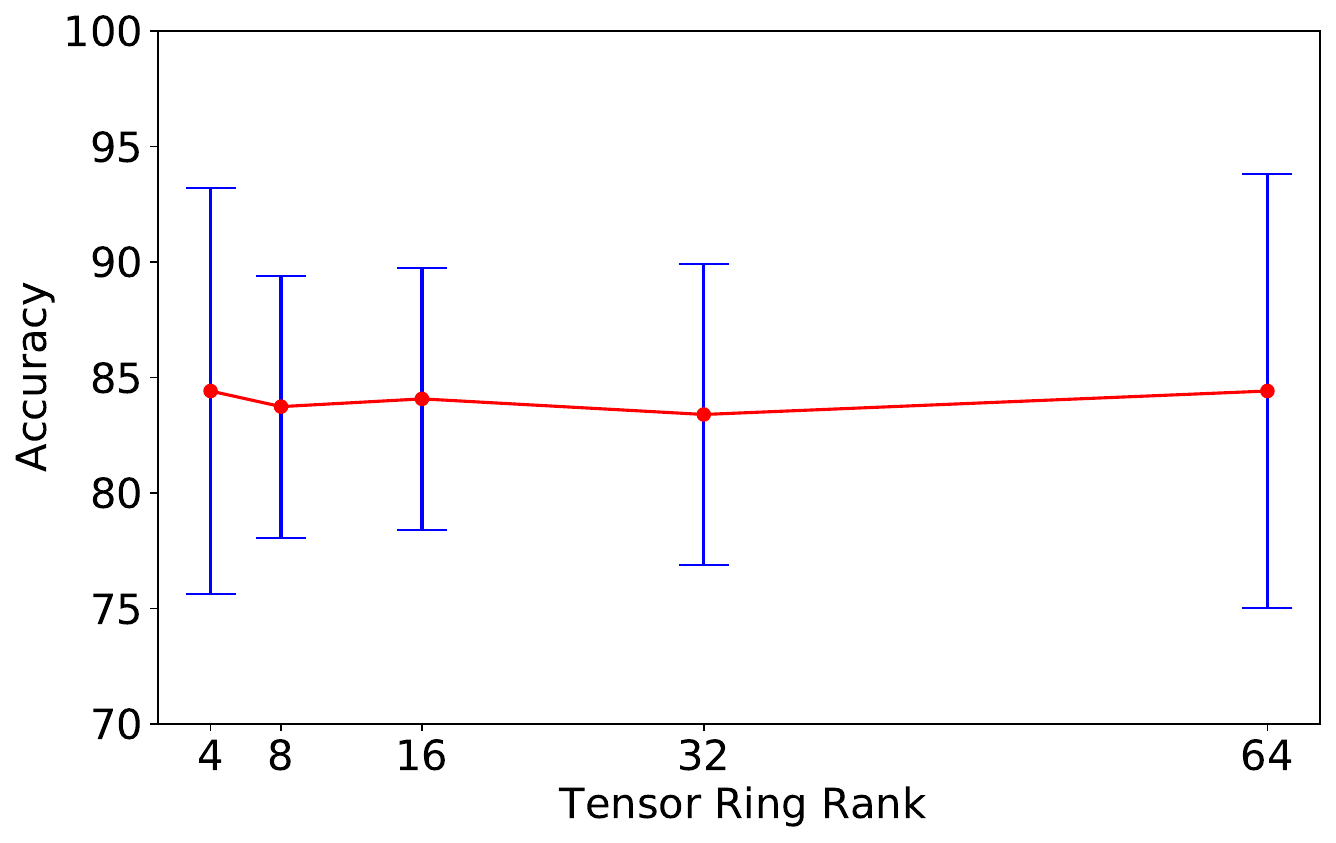}\label{fig:mnist2}}
	\hfill
	\subfloat{\includegraphics[keepaspectratio,width=0.5\textwidth]{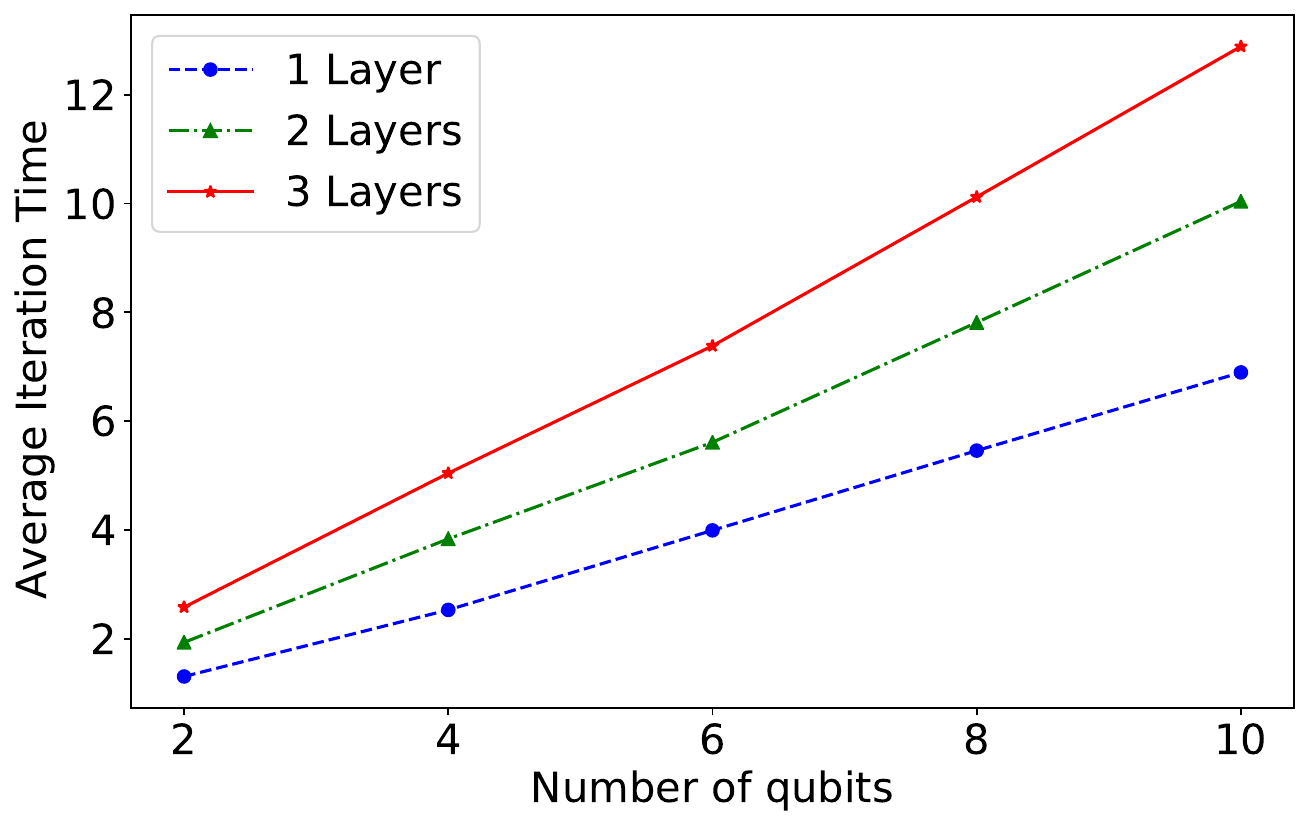}\label{fig:mnist3}}

\caption{(a) Performance of Tensor Ring VQC with respect to varying rank for a 4 qubits, 2 layer circuit on reduced MNIST data (b) Average iteration times with varying number of qubits and circuit depth with reduced MNIST data}%
\label{fig:MNIST}%
\end{figure*}

\subsection{Binary Classification: MNIST} \label{sec:4.2}
 We perform a binary classification on MNIST dataset with classes $3$ and $7$ (among the ten classes from 0 to 9). We choose 230 samples with labels $3$ and $7$ (referred to as reduced MNIST in this manuscript) to be split into training and test datasets to facilitate faster training. We have also chosen a smaller dataset for effective comparison against Qiskit's algorithm which can prove harder to train over larger datasets. MNIST dataset consists of images of size $28\times28$, which is reduced to a smaller feature vector using Principal Component Analysis (PCA). We generate a feature space of dimension 4 and 8 for multiple experiments. The circuit structure uses the same layer as the experiment in IRIS dataset, however multiple experiments are conducted by repeating these layers multiple times as seen in Table \ref{tab:results}. The initially prepared qubit state $\ket{0000}$  for a 4 dimension feature space is decomposed into a tensor ring as described before. We select the final measurements $\ket{0000}$ and $\ket{1111}$  as output values for 2 classes (The same method, scaled appropriately is used to encode outputs for experiments with larger number of qubits). We further apply softmax function to selected measurements to obtain class probability and compute the cross-entropy loss and backpropagate to obtain gradients. The VQC is trained with an Adam optimizer with a learning rate of $0.01$ and a batch size of $4$ for $50$ epochs. The accuracy from the simulations are reported in the Table \ref{tab:results}.

 \subsection{Discussion}\label{sec:4.3}
 The overall results are summarized in Table \ref{tab:results}. We found $4$ and $8$ to be the optimal number of qubits for performance for the given datasets. Increasing  further to $16$ qubits led to a drastic reduction in the accuracy  for both the proposed method and the benchmark possibly due to overparametrization. In most cases, the tensor ring parametrized VQC performs comparably or outperforms the implementation of MPS-VQC on Qiskit  with full quantum state information owing to the cyclic structure and the additional non-linearity induced by the truncated singular value decomposition over the two qubit gate transformations. In order to observe the effect of tensor ring rank on the performance of VQCs, a series of experiments using various ranks are conducted with 4 qubit and 2 layer circuit architectures  with reduced MNIST data whose results are illustrated in Figure \ref{fig:mnist2}. It is to be noted that the accuracy does not change significantly with an increasing TR rank indicating that a considerable compression can be achieved using smaller ranks  without losing out on the performance. We also plot the average iteration times (an iteration involves a forward pass and parameter optimization for a given batch size) with respect to the number of qubits and circuit layers in \ref{fig:mnist3} to demonstrate the linear complexity of the TRVQCs.
 
We note that we never store the full quantum state, unlike in standard quantum circuit. So, at each step of the operation, we only store the $N$ tensors, giving an overall storage complexity of $O(N\chi^2)$, where the tensor-ring rank $\chi$ is a hyper-parameter. For computational complexity, each computation of single or two-qubit gate in the tensor ring VQC is $O(1)$, and thus the overall complexity of the forward pass is $O(Nd)$, or the size of the circuit/number of gates. Similarly, each backpropagation round is the same complexity. Thus, each iteration of training has complexity  $O(Nd)$ making the overall approach scalable in the number of qubits. In order to evaluate the final amplitudes which we will use to map to class probabilities, we contract the transformed tensor ring to obtain the final quantum state. It is to be noted that although the computational complexity of tensor ring contractions is $O(N\chi^3)$, the storage complexity would explode to $O(2^N)$ upon contraction since it results in the full quantum state.  Although not utilized in this work, more efficient methods can be used to contract the tensor network as discussed in Ferris et al. \cite{ferris2012perfect} which uses Monte Carlo simulations to compute the expectation, reducing the computational complexity to  $O(N\chi^2)$.

 \section{Conclusion} \label{sec:5}
 In this work, we propose a new technique to train Variational Quantum Circuits for supervised learning on classical simulators using an approximate tensor ring representation of the quantum state. The computational time of the circuit evaluation grows linearly in the number of qubits, $N$, and the depth of the circuit, $d$, with the proposed method, as compared to the exponential growth in the simulation of ideal quantum circuits. We train the Tensor Ring Variational Quantum Circuits using the proposed approach on Iris and a subset of MNIST dataset. The performance is significantly better than the implementation of VQCs on classical simulators using full state MPS approximation in case of MNIST data. This demonstrates that the proposed architecture using TR representation could possibly outperform methods that don't use approximation. Given the linear complexity as demonstrated by the iteration times from the simulations, scalability in number of qubits and layers can be achieved more efficiently using the proposed approximation which helps in larger scale applications along with a potential improvement in performance. It is also to be noted that the proposed approach can be used to train VQCs on classical devices before implementing on quantum simulators potentially reducing or eliminating any training on the quantum simulators.

We note that the quantum circuit operations (except measurement) are linear in the Hilbert space, while the added approximation of two-qubit gates bring non-linearity in the circuit, similar to Rectified Linear Unit (ReLU) or soft-max in the classical neural networks. More rigorous work is required to identify theoretical guarantees for improved performance with the addition of non-linearity in VQCs.  One of the major challenges in training large scale quantum circuits is the barren plateau effect \cite{mcclean2018barren} which demands initialization exploration to achieve comparable results. Barren plateuas also limit the performance of the algorithms implemented on tensor network based classical simulators, including the proposed method wanting for methods that address the issue. As for future work along this direction, building up on the limitations discussed, barren plateau mitigation methods specific to the tensor ring approximations must be studied for scalability. Another potential application of the proposed work would be as an initialization technique for QML algorithms before being implemented on quantum processors. It would also be interesting to see the performance of proposed approach replacing neural networks in tasks like Reinforcement Learning or Generative Adversarial Networks, etc.

\appendix

\section{Matrix Representations}
The matrix representation of some of the commonly used gates in the manuscript are listed below:
$$
R_x(\theta) = 
\begin{bmatrix}
cos(\theta/2) &  -isin(\theta/2)\\
-isin(\theta/2) & cos(\theta/2)
\end{bmatrix} 
$$

$$
R_y(\theta) = 
\begin{bmatrix}
cos(\theta/2) &  -sin(\theta/2)\\
sin(\theta/2) & cos(\theta/2)
\end{bmatrix}
$$

$$
R_z(\theta) = 
\begin{bmatrix}
e^{-i\theta/2} &  0\\
0 & e^{i\theta/2}
\end{bmatrix}
$$

$$
H =\frac{1}{\sqrt{2}}
\begin{bmatrix}
 1&  1\\
1 & -1
\end{bmatrix}
$$

$$
CNOT = 
\begin{bmatrix}
1 & 0 & 0 & 0\\
0 & 1 & 0 & 0\\
0 & 0 & 0 & 1\\
0 & 0 & 1 & 0
\end{bmatrix}
$$

$$
R(\alpha, \beta, \gamma) = 
\begin{bmatrix}
cos(\alpha/2) & -e^{i\gamma}sin(\alpha/2) \\
e^{i\beta}sin(\alpha/2) & e^{i\beta + i\gamma}cos(\alpha/2)
\end{bmatrix}
$$

\bibliographystyle{elsarticle-num} 
\bibliography{refs}

\begin{thebibliography}{10}
\expandafter\ifx\csname url\endcsname\relax
  \def\url#1{\texttt{#1}}\fi
\expandafter\ifx\csname urlprefix\endcsname\relax\def\urlprefix{URL }\fi
\expandafter\ifx\csname href\endcsname\relax
  \def\href#1#2{#2} \def\path#1{#1}\fi

\bibitem{10.1145/237814.237866}
L.~K. Grover, \href{https://doi.org/10.1145/237814.237866}{A fast quantum
  mechanical algorithm for database search}, in: Proceedings of the
  Twenty-Eighth Annual ACM Symposium on Theory of Computing, STOC '96,
  Association for Computing Machinery, New York, NY, USA, 1996, p. 212–219.
\newblock \href {https://doi.org/10.1145/237814.237866}
  {\path{doi:10.1145/237814.237866}}.
\newline\urlprefix\url{https://doi.org/10.1145/237814.237866}

\bibitem{shor1999polynomial}
P.~W. Shor, Polynomial-time algorithms for prime factorization and discrete
  logarithms on a quantum computer, SIAM review 41~(2) (1999) 303--332.

\bibitem{cerezo2021variational}
M.~Cerezo, A.~Arrasmith, R.~Babbush, S.~C. Benjamin, S.~Endo, K.~Fujii, J.~R.
  McClean, K.~Mitarai, X.~Yuan, L.~Cincio, et~al., Variational quantum
  algorithms, Nature Reviews Physics (2021) 1--20.

\bibitem{zhou2020quantum}
L.~Zhou, S.-T. Wang, S.~Choi, H.~Pichler, M.~D. Lukin, Quantum approximate
  optimization algorithm: Performance, mechanism, and implementation on
  near-term devices, Physical Review X 10~(2) (2020) 021067.

\bibitem{farhi2014quantum}
E.~Farhi, J.~Goldstone, S.~Gutmann, A quantum approximate optimization
  algorithm, arXiv preprint arXiv:1411.4028 (2014).

\bibitem{mcclean2016theory}
J.~R. McClean, J.~Romero, R.~Babbush, A.~Aspuru-Guzik, The theory of
  variational hybrid quantum-classical algorithms, New Journal of Physics
  18~(2) (2016) 023023.

\bibitem{mitarai2018quantum}
K.~Mitarai, M.~Negoro, M.~Kitagawa, K.~Fujii, Quantum circuit learning,
  Physical Review A 98~(3) (2018) 032309.

\bibitem{chen2020quantum}
S.~Y.-C. Chen, S.~Yoo, Y.-L.~L. Fang, Quantum long short-term memory, arXiv
  preprint arXiv:2009.01783 (2020).

\bibitem{cong2019quantum}
I.~Cong, S.~Choi, M.~D. Lukin, Quantum convolutional neural networks, Nature
  Physics 15~(12) (2019) 1273--1278.

\bibitem{liu2021hybrid}
J.~Liu, K.~H. Lim, K.~L. Wood, W.~Huang, C.~Guo, H.-L. Huang, Hybrid
  quantum-classical convolutional neural networks, Science China Physics,
  Mechanics \& Astronomy 64~(9) (2021) 1--8.

\bibitem{Du_2020}
Y.~Du, M.-H. Hsieh, T.~Liu, D.~Tao,
  \href{http://dx.doi.org/10.1103/PhysRevResearch.2.033125}{Expressive power of
  parametrized quantum circuits}, Physical Review Research 2~(3) (Jul 2020).
\newblock \href {https://doi.org/10.1103/physrevresearch.2.033125}
  {\path{doi:10.1103/physrevresearch.2.033125}}.
\newline\urlprefix\url{http://dx.doi.org/10.1103/PhysRevResearch.2.033125}

\bibitem{schollwock2011density}
U.~Schollw{\"o}ck, The density-matrix renormalization group in the age of
  matrix product states, Annals of physics 326~(1) (2011) 96--192.

\bibitem{PhysRevX.10.041038}
Y.~Zhou, E.~M. Stoudenmire, X.~Waintal,
  \href{https://link.aps.org/doi/10.1103/PhysRevX.10.041038}{What limits the
  simulation of quantum computers?}, Phys. Rev. X 10 (2020) 041038.
\newblock \href {https://doi.org/10.1103/PhysRevX.10.041038}
  {\path{doi:10.1103/PhysRevX.10.041038}}.
\newline\urlprefix\url{https://link.aps.org/doi/10.1103/PhysRevX.10.041038}

\bibitem{wang2017efficient}
W.~Wang, V.~Aggarwal, S.~Aeron, Efficient low rank tensor ring completion, in:
  Proceedings of the IEEE International Conference on Computer Vision, 2017,
  pp. 5697--5705.

\bibitem{Wang_2018_CVPR}
W.~Wang, Y.~Sun, B.~Eriksson, W.~Wang, V.~Aggarwal, Wide compression: Tensor
  ring nets, in: Proceedings of the IEEE Conference on Computer Vision and
  Pattern Recognition (CVPR), 2018.

\bibitem{malik2021sampling}
O.~A. Malik, S.~Becker, A sampling-based method for tensor ring decomposition,
  in: International Conference on Machine Learning, PMLR, 2021, pp. 7400--7411.

\bibitem{vidal2003efficient}
G.~Vidal, Efficient classical simulation of slightly entangled quantum
  computations, Physical review letters 91~(14) (2003) 147902.

\bibitem{markov2008simulating}
I.~L. Markov, Y.~Shi, Simulating quantum computation by contracting tensor
  networks, SIAM Journal on Computing 38~(3) (2008) 963--981.

\bibitem{white1992density}
S.~R. White, Density matrix formulation for quantum renormalization groups,
  Physical review letters 69~(19) (1992) 2863.

\bibitem{verstraete2008matrix}
F.~Verstraete, V.~Murg, J.~I. Cirac, Matrix product states, projected entangled
  pair states, and variational renormalization group methods for quantum spin
  systems, Advances in physics 57~(2) (2008) 143--224.

\bibitem{vidal2007entanglement}
G.~Vidal, Entanglement renormalization, Physical review letters 99~(22) (2007)
  220405.

\bibitem{huggins2019towards}
W.~Huggins, P.~Patil, B.~Mitchell, K.~B. Whaley, E.~M. Stoudenmire, Towards
  quantum machine learning with tensor networks, Quantum Science and technology
  4~(2) (2019) 024001.

\bibitem{cheng2021supervised}
S.~Cheng, L.~Wang, P.~Zhang, Supervised learning with projected entangled pair
  states, Physical Review B 103~(12) (2021) 125117.

\bibitem{efthymiou2019tensornetwork}
S.~Efthymiou, J.~Hidary, S.~Leichenauer, Tensornetwork for machine learning,
  arXiv preprint arXiv:1906.06329 (2019).

\bibitem{stoudenmire2016supervised}
E.~Stoudenmire, D.~J. Schwab, Supervised learning with tensor networks,
  Advances in Neural Information Processing Systems 29 (2016).

\bibitem{orus2014practical}
R.~Or{\'u}s, A practical introduction to tensor networks: Matrix product states
  and projected entangled pair states, Annals of physics 349 (2014) 117--158.

\bibitem{watabe2019quantum}
M.~Watabe, K.~Shiba, M.~Sogabe, K.~Sakamoto, T.~Sogabe, Quantum circuit
  parameters learning with gradient descent using backpropagation, arXiv
  preprint arXiv:1910.14266 (2019).

\bibitem{li2009efficient}
C.~H. Li, S.~C. Park, An efficient document classification model using an
  improved back propagation neural network and singular value decomposition,
  Expert Systems with Applications 36~(2) (2009) 3208--3215.

\bibitem{du2020quantum}
Y.~Du, T.~Huang, S.~You, M.-H. Hsieh, D.~Tao, Quantum circuit architecture
  search: error mitigation and trainability enhancement for variational quantum
  solvers, arXiv preprint arXiv:2010.10217 (2020).

\bibitem{ferris2012perfect}
A.~J. Ferris, G.~Vidal, Perfect sampling with unitary tensor networks, Physical
  Review B 85~(16) (2012) 165146.

\bibitem{mcclean2018barren}
J.~R. McClean, S.~Boixo, V.~N. Smelyanskiy, R.~Babbush, H.~Neven, Barren
  plateaus in quantum neural network training landscapes, Nature communications
  9~(1) (2018) 1--6.

\end{thebibliography}

\end{document}